\documentclass[]{elsarticle}
\usepackage{lineno,hyperref}
\modulolinenumbers[5]
\usepackage{amsmath,empheq}
\usepackage{amsfonts}
\usepackage{amsthm}
\usepackage{amssymb}
\usepackage{graphicx}
\usepackage{hyperref}
\usepackage{soul}
\usepackage[normalem]{ulem}
\usepackage{textcomp}
\usepackage{commath} 
\usepackage{xcolor}
\usepackage{tikz}
\usepackage{tikz-cd}
\usetikzlibrary{%
  matrix,%
  calc,%
  arrows%
}

\usepackage{geometry}

	\newcommand{\ncd}{\newcommand}
	\ncd{\mrm}    {\mathrm}
	\ncd{\beq} {\begin{equation}}
	\ncd{\eeq} {\end{equation}}
	\ncd{\nn}{\nonumber}

	\def\d{{\rm d}}

	\def\basis[#1]{\frac{\partial}{\partial #1}}
	\def\dt[#1]{\frac{\d}{\d #1}}

\begin{document}

\begin{frontmatter}

\title{The geometry of  induced electromagnetic fields in  moving media}

		\author[cslm_address]{C. S. Lopez-Monsalvo}
		\ead{cslopezmo@conacyt.mx}
		
		\author[dgp_address1,dgp_address2]{D. Garcia-Pelaez}
		\ead{dgarciap@up.edu.mx}

		\author[dgp_address1]{A. Rubio-Ponce}
		\ead{arp@azc.uam.mx}
		
		\author[dgp_address1]{R. Escarela-Perez}
		\ead{rep@azc.uam.mx}

		\address[cslm_address]{Conacyt-Universidad Aut\'onoma Metropolitana Azcapotzalco
		Avenida San Pablo Xalpa 180, Azcapotzalco, Reynosa Tamaulipas, 02200 Ciudad de 
		M\'exico, M\'exico}
		
		\address[dgp_address1]{Universidad Aut\'onoma Metropolitana Azcapotzalco
		Avenida San Pablo Xalpa 180, Azcapotzalco, Reynosa Tamaulipas, 02200 Ciudad de 
		M\'exico, M\'exico}
		
		\address[dgp_address2]{Universidad Panamericana, Tecoyotitla 366. Col. Ex Hacienda Guadalupe Chimalistac,
		C.P. 01050 Ciudad de Mexico, Mexico}

\begin{abstract}
In this manuscript we provide a fully geometric formulation for the induced electromagnetic fields and their corresponding constitutive relations  in moving media. To this end, we present the reader with a brief geometric summary  to show  how vector calculus electromagnetic theory is embedded in the more general language of differential forms. Then, we consider the class of \emph{metric} constitutive relations describing  the medium in which electromagnetic fields propagate. We explicitly obtain the components of the induced  fields in a moving medium,  as seen in the the lab \emph{rest} frame. This allows us to read the expressions for the permitivity, permeability and magnetoelectric matrices for the moving medium which, in turn, can be interpreted as a different physical material from the lab point of view. 


\end{abstract}


\end{frontmatter}

\section{Introduction}

It has been since the early days of General Relativity that we have seen that the ``influence of matter on electromagnetic phenomena is equivalent to the influence of a gravitational field'' \cite{gordon1923lichtfortpflanzung, EhlersJ2012GFFandLP}. That is, in the same manner light rays obey Fermat's principle while propagating across media, in General Relativity light follows null geodesics on a curved spacetime. Thus, it has been argued that spacetime acts like a medium with a particular refractive index, where all  the information is encoded in its metric tensor \cite{de1971gravitational,de2002analogue,novello2003analogue,belgiorno2011dielectric}. Thus, we can reverse the argument and note that optical media can be treated geometrically by means of differentiable manifolds where light follows the corresponding curvature \cite{hehl2012foundations,asenjo2017differential}. Such intuition has been  exploited in the recent development of material science and engineering \cite{leonhardt2006general,zheludev2010road,asenjo2017differential,schuster2017effective,thompson2018covariant,schuster2019electromagnetic, PendryJ06-CEF}. 

The formulation of a field theory in the language of differential geometry has been thoroughly exploited during the last century. However, most of the work done so far has been developed to pursue goals in fundamental areas of theoretical physics \cite{misner1957classical,misner1973gravitation}. It has only been in recent times that these tools have begun to be used in more applied areas \cite{baldomir1996geometry,bossavit1999computational,leonhardt2010geometry,stenvall2013manifolds}. For instance, in material science, all information regarding the macroscopic response of a medium to electromagnetic stimuli is encoded in its constitutive tensor, which has been related to a metric or a curvature of the geometric space represented by the medium. \cite{asenjo2017differential, gordon1923lichtfortpflanzung}. 

The constitutive relations are usually expressed in terms of the permittivity, permeability and magnetoelectric matrices.  These, however, are usually written and interpreted in terms of a single set of coordinates within the vector calculus formulation of electromagnetism.  Thus, one of our aims is to explicitly bridge such a formulation with the coordinate and frame independent differential form language. We do this constructively, exhibiting the fact that Maxwell's equations are conservation laws in spacetime while constitutive relations are maps linking the differential forms associated with these conservation laws.

It is of pedagogical value to  see how vector calculus electromagnetic theory is embedded in the more general differential form language. Such details are, more often than not, omitted in the modern literature based on differential geometry. Therefore, in section \ref{sec.vector} we recall the traditional formulation of electromagnetic theory starting from the integral form of Maxwell's equations in domains of $\mathbb{R}^3$ followed, in section \ref{sec.diffforms}, of their generalization to a general differentiable manifold $\mathcal{M}$.  Notoriously, in  formulating Maxwell's equations, there is no need to equip the manifold $\mathcal{M}$ with a metric tensor. However, it is clear that there is no link between the sources and the fields. Such a link could take various guises, yet it is specially convenient if it is through an intrinsic geometric structure associated with the manifold. In this way,  one can guarantee that the formulation is independent of the choice of coordinates and observers.  Moreover, it comes as an additional postulate that such a structure contains all the relevant macroscopic electromagnetic information of the material where the fields are propagating \cite{hehl2012foundations}.

Here, we adhere to the view that different materials are described by different geometries. That is, we assume that  constitutive relations are expressed in terms of the \emph{Hodge} dual operator associated with each material metric tensor.   Therefore, we consider a metric for the ambient space and a metric for the medium. It is worth noting that this is not the most general way to \emph{geometrize}  constitutive relations but  it is the certainly one of the simplest. As a result, we obtain a general and coordinate free expression to explicitly compute the components of the induced electromagnetic vector fields  as seen by an arbitrary observer. This is done in section \ref{sec.metrics}.


In section \ref{sec.movmedia} we consider the effect of external electromagnetic fields on a moving medium which is homogeneous and isotropic when it is at rest in the lab frame. The corresponding induced fields  are described by a metric tensor adapted to the motion of the medium. Such motion, defines a coordinate transformation  which maps the material metric into its moving version. 

Here, we study various types of transformations. First, we consider a medium moving at constant velocity  with respect to the static laboratory frame. Then, we analyze the case of  non-inertial motion. In particular, when the medium is undergoing uniform acceleration and the case when it is rotating. In all cases we make both analysis, Galilean and relativistic \footnote{Galilean and relativistic analysis are so called in terms of the coordinate transformations of the moving media. In the Galilean case, we do not intend to do a low velocity limit.}. Interestingly, the transformation describing rotating objects consistent with the principles of special relativity remain a timely subject \cite{gron2004space,rizzi2013relativity,gourgoulhon2016special}.  We obtain the corresponding metric for the moving medium and explicitly obtain the permittivity, permeability and magnetoelectric matrices.

As noted originally by R$\ddot{\rm o}$ntgen \cite{rontgen1888ueber}, a  medium immersed  in a purely electric field, as measured by a static observer,  appears to be magnetized when it moves with respect the static frame. Similarly, there is the corresponding apparent polarization when we replace the electric by a magnetic field. In all cases, the resulting constitutive relations for the moving medium couple the electric and magnetic fields . This is known as the magnetoelectric effect (see \cite{fiebig2005revival} and reference therein for a timely description) and it has become a very active research area in material science, e.g. due to the possibility of controlling the magnetization of a ferromagnet  rotated by means of purely electric field \cite{spaldin2019advances}. In this work, we explicitly extract the magnetoelectric matrices of a simple medium for each type of \emph{elementary} motion. Moreover, due to the generality of the geometric framework, the same analysis can be readily exported to more complicated materias, described by curved geometries, in arbitrary motion. This is done in section \ref{sec.nontrivial}, where we consider a non-trivial medium associated with a curved metric and a non-inertial transformation.

Finally, in section \ref{sec.concl} we provide some closing remarks and provide some further directions for exploration.

Throughout the manuscript, we decline the use of the Einstein sum convention and refrain of using a designated letter for the speed of light in vacuum as well as in the medium. This served as bookkeeping of  all the geometric factors involved in the transformations. Thus, albeit our expressions are slightly longer, they provide a clearer notion of scales and units.   

%

\section{Vector calculus electromagnetism}
\label{sec.vector}

The empirical character of electromagnetism lies on the  fact that in nature there is a distinguished property of matter that  certain objects posses and which can be perceived by means of its motion and interaction. Such property is  \emph{observed} to be conserved and it is called electric charge. Accordingly,  we infer the existence of a field responsible for the inertial change of the charges and, in turn, as charges move around a new field configuration arises.  The field itself obeys its own conservation law and this lead us to a dynamical theory of fields and charges. This is expressed as a series of observed relations between  fields and  sources, namely,
	\begin{align}
	\label{eq.max1}
	\oint_{\partial \Omega} \vec B \cdot \hat{n} \ \d s & = 0,\\
	\label{eq.max2}
	\oint_{\partial \Sigma} \vec E \cdot \d \vec \ell & = - \frac{\d}{\d t} \int_\Sigma \vec B \cdot \hat{n} \ \d s,\\
	\label{eq.max3}
	\oint_{\partial \Omega} \vec D \cdot \hat{n} \ \d s & = \int_\Omega \rho_{\rm ext} \d v
	\end{align}
and
	\beq
	\label{eq.max4}
	\oint_{\partial \Sigma} \vec H \cdot \d \vec \ell  = \frac{\d}{\d t}\int_\Sigma \vec D \cdot \hat{n}\  \d s + \int_\Sigma \vec j_{\rm ext} \cdot \hat{n} \ \d s.
	\eeq
Here, we refer to $\vec B$ and $\vec E$ as the \emph{fundamental}  magnetic and electric fields, respectively, while $\vec H$ and $\vec D$ represent the corresponding \emph{induced} fields in a given medium. The terms $\rho_{\rm ext}$ and $\vec j_{\rm ext}$ are the \emph{external} electric charge density and current density  flux, respectively and represent the sources of the fields. Notice that the induced fields are the ones linked to the sources while the fundamental fields seem to be independent.  The symbol $\partial$ is known as the \emph{boundary} operator, in this case acting on domains of $\mathbb{R}^3$. Thus $\partial \Omega$ is the 2-dimensional boundary of a 3-dimensional open region $\Omega\subset \mathbb{R}^3$, while $\partial \Sigma$ is the 1-dimensional curve bounding an open surface $\Sigma\subset \mathbb{R}^3$.

The passing from the global representation to the local expressions of Maxwell's equations is a straightforward application of the vector calculus integral theorems. Thus it follows that Maxwell's equations, in their local form,  can be separated into  the homogeneous 
	\begin{align}
	\label{eq.maxGM}
	\nabla \cdot  \vec B & = 0, \\
	\label{eq.maxFar}
	\nabla \times  \vec E +\frac{\partial}{\partial t}  \vec B & = 0,
	\end{align}
and  in-homogenous 
	\begin{align}
	\label{eq.EGauss}
	\nabla \cdot \vec D & = \rho_{\rm ext}  \\
	\label{eq.ampere}
	\nabla \times \vec H - \frac{\partial}{\partial t} \vec D & =  \vec j_{\rm ext},
	\end{align}
pairs of equations.


 An immediate consequence of this  is a continuity equation for the sources. That is, applying the divergence operator  and substituting \eqref{eq.EGauss} into \eqref{eq.ampere} it follows that
	\beq
	\label{eq.cont}
	\frac{\partial}{\partial t} \rho_{\rm ext} + \nabla \cdot \vec j_{\rm ext} = 0.
	\eeq 
Note that this conservation law only refers to the external charges and currents. In addition to the external sources, each medium is characterized by a response function to the externally applied fields, implying the appearance of induced charges and currents within the materials. Therefore, assuming the conservation of \emph{total} charge entails that  the induced charges and currents must be conserved independently and hence, there is no interchange between external and induced charges. Thus, postulating Maxwell's equations \eqref{eq.maxGM} -- \eqref{eq.ampere} together with the conservation of \emph{total} charge
	\beq
	\label{eq.conttot}
	\frac{\partial}{\partial t} \rho + \nabla \cdot \vec j = 0,
	\eeq	
implies the conservation law for the \emph{induced} charge and current densities
	\beq
	\label{eq.cont2}
	\frac{\partial}{\partial t} \rho_{\rm ind} + \nabla \cdot \vec j_{\rm ind} = 0.
	\eeq
Here,
	\beq
	\rho = \rho_{\rm ext} + \rho_{\rm ind} \quad \text{and} \quad \vec j = \vec j_{\rm ext} + \vec j_{\rm ind}.
	\eeq

The fundamental problem of any field theory consists in determining the fields from the known sources and some \emph{a priori} information about the fields in a certain region of space and time. In the case of electromagnetism  we have to determine the electric field $\vec E$ and the magnetic flux $\vec B$ from the known functions of space and time $\rho_{\rm ext}$ and $\vec j_{\rm ext}$ together with a set of prescribed boundary and initial conditions. Formulated in this manner, the problem is incomplete, since there is no link between the homogeneous equations \eqref{eq.maxGM} and \eqref{eq.maxFar} and the source equations \eqref{eq.EGauss} and \eqref{eq.ampere}. That is, an extra set of equations known as the \emph{constitutive relations of the medium} has to be imposed.

The constitutive relations incorporate information about the medium response to the  stimuli produced by  external fields. In general, these are expressed in terms of a convolution averaging the field effect over the entire space occupied by the medium through the material's complete history. In the simplest scenario, these can be expressed as the  linear transformations  \cite{DahlM04-CGE}   
	\beq
	\label{eq.const1}
	\left(
	\begin{array}{c}
		\vec D \\
		\vec H
	\end{array} \right) = \left(
					\begin{array}{cc}
						\bar\varepsilon 	& \bar\zeta \\
						\bar\chi	& \bar\mu^{-1}
					\end{array} \right)	\left(
								 	\begin{array}{c}
										\vec E \\
										\vec B
									\end{array}\right),
	\eeq
where $\bar\varepsilon$ and $\bar\mu^{-1}$ are the $3\times 3$ permittivity and (inverse) permeability matrices, respectively, and $\bar\zeta$ and $\bar\chi$\footnote{Regarding this work, all media will be supposed dielectric, for which $\bar\zeta$ and $\bar\chi$ are always real. } are the so called magnetoelectric matrices \cite{fiebig2005revival,schuster2017effective}.

In the following sections we present electromagnetic theory in the language of differential forms and Riemannian geometry. There are numerous references on this subject. For the details concerning definitions and operational tools from a physical point of view the standard texts \cite{nash1988topology,nakahara2003geometry} are recommended. For more formal details on the mathematical side, we use the conventions of \cite{kobayashi1963foundations}. For the applications  of differential geometry in the science and engineering of electromagnetic fields  we urge the reader to consult \cite{baldomir1996geometry,gross2004electromagnetic, hehl2012foundations}.

\section{Electromagnetism in differential forms}
\label{sec.diffforms}

Maxwell's equations are empirical postulates requiring the conservation of certain quantities. Conservation laws are best understood in their integral form. One usually considers a flux crossing the \emph{boundary} of a certain region and imposes its conservation. Then, using Stokes' theorem and  the arbitrariness of the region of interest one observes that demanding the conservation of the flux is equivalent to requiring its correspondence to a closed differential form. Schematically
	\beq
	0 \overset{!}{=} \oint_{\partial \Omega} J = \int_{\Omega} \d J \quad \forall \Omega \subset \mathcal{M} \implies \d J =0.
	\eeq 
Here, $J$ is  $p$-form  (with $0<p< {\rm dim}\mathcal{M}$) representing a $p$-flux, $\Omega$ is a $p+1$ dimensional region of $\mathcal{M}$ with a  $p$-dimensional boundary $\partial \Omega$, e.g. a 2-dimensional surface bounded by a closed curve, a 3-dimensional volume bounded by a closed surface or, analogously, a 4-dimensional region bounded by a closed volume. Also, we use the symbol $\overset{!}{=}$ to express the empirical imposition of such equality.  Therefore, Maxwell's equations are postulated as the conservation laws for a 2-form $F$, that is
	\beq
	\label{eq.post1}
	\oint_{\partial \Omega^3} F \overset{!}{=} 0 \quad \forall \Omega^3 \subset \mathcal{M} \implies \d F \overset{!}{=}0,
	\eeq
and an $n-1$-form $j$, i.e.
	\beq
	\label{eq.post2}
	\oint_{\partial \Omega^n} j \overset{!}{=} 0 \quad \forall \Omega^n \subset \mathcal{M} \implies \d j \overset{!}{=}0.
	\eeq
These statements are empirical postulates and are completely general, i.e. they are coordinate free, observer independent and require no further structure other than differentiability of $\mathcal{M}$. The former, states the conservation of the total electromagnetic flux whilst, the latter, the conservation of the total charge. Therefore, Maxwell's equations \eqref{eq.maxGM} -- \eqref{eq.ampere} can be written in terms of differential forms on a 4-dimensional manifold $\mathcal{M}$ as
	\beq
	\label{eq.MEdf1}
	\d F = 0,
	\eeq
and
	\beq
	\label{eq.MEdf2}
	 \d G - j_{\rm ext} =0
	\eeq
where, as before, we have the homogeneous and source equations. Here $F$ and $G$ are 2-forms representing the $(\vec E,\vec B)$ and $(\vec D, \vec H)$ fields, respectively, while $j_{\rm ext}$ is a 3-form representing the free sources $(\rho_{\rm ext},\vec j_{\rm ext})$ and corresponds to the part of the total current density three form  which is not induced by the fields in the medium
	\beq
	j_{\rm ext} = j - j_{\rm ind}.
	\eeq

Similar to equations \eqref{eq.maxGM} -- \eqref{eq.ampere}, equations \eqref{eq.MEdf1} and \eqref{eq.MEdf2} are coordinate independent, that is, they remain valid regardless of the choice of local coordinates for $\mathcal{M}$. Thus, to convince ourselves  that, indeed, equations \eqref{eq.MEdf1} and \eqref{eq.MEdf2} are equivalent to equations \eqref{eq.maxGM} -- \eqref{eq.ampere}, let us work in a cartesian coordinate system $(x^1,x^2,x^3,x^4) = (x,y,z,t)$ for an open set of $\mathcal{M}$. 

Let
	\beq
	\label{eq.faraday}
	F = B + E \wedge \d t,
	\eeq
	\beq
	\label{eq.fardaydual}
	 G = - D + H \wedge \d t
	\eeq
and
	\beq
	\label{eq.current}
	j_{\rm ext} = -\rho_{\rm ext}  + j^{(3)}_{\rm ext}  \wedge \d t.
	\eeq
Here, the fields $E$ and $H$ are the 1-forms whose components are equal to their vectorial counterparts, i.e.
	\beq
	E = \sum_{i=1}^3 E_i \d x^i = E_x \d x + E_y \d y + E_z \d z
	\eeq
and
	\beq
	H = \sum_{i=1}^3 H_i \d x^i = H_x \d x + H_y \d y + H_z \d z
	\eeq
whilst the fluxes $B$, $D$ and $j^{(3)}_{\rm ext}$ are the 2-forms
	\beq
	\label{eq.Bform}
	B = \sum_{\overset{i,j=1}{i\neq j}}^3 B_{ij} \d x^i \wedge \d x^j,
	\eeq
	\beq
	D = \sum_{\overset{i,j=1}{i\neq j}}^3 D_{ij} \d x^i \wedge \d x^j,
	\eeq
and 
	\beq
	j^{(3)}_{\rm ext} = \sum_{\overset{i,j=1}{i\neq j}}^3 j_{ij} \d x^i \wedge \d x^j,
	\eeq
where $B_{ij}$ (resp. $D_{ij}$ and $j_{ij}$) represents the magnetic (resp. electric and external current density) flux crossing the infinitesimal oriented area element  $\d x^i \wedge \d x^j$ \cite{WarnickK-2014} i.e.  the component of $\vec B \in \mathbb{R}^3$ (resp. $\vec D$ and $\vec j_{\rm ext}$) orthogonal to the space generated by $\hat{e}_{(i)}$ and $\hat{e}_{(j)}$, namely
	\beq
	\label{eq.fluxcomp}
	B_{ij} = \vec B \cdot \hat{e}_{(k)} = B_{k} \quad \text{with} \quad   \hat{e}_{(i)} \cdot  \hat{e}_{(j)} =  \hat{e}_{(i)} \cdot  \hat{e}_{(k)} = \hat{e}_{(j)} \cdot \hat{e}_{(k)}= 0
	\eeq
(resp.  $D_{ij} = \vec D \cdot \hat k = D_k$ and  $j_{ij} = \vec j_{\rm ext} \cdot \hat k = {j_{\rm ext}}_k$) and, finally, $\rho_{\rm ext}$ is the external charge density 3-form
	\beq
	\rho_{\rm ext} = \rho_{\rm ext}\  \d x \wedge \d y \wedge \d z.
	\eeq
Here, we are using the Cartesian dot product merely to illustrate how the components of the  vector fields in $\mathbb{R}^3$ are related to those of their corresponding differential forms. It is not an additional structure over the manifold $\mathcal{M}$.
 
It is a straightforward algebraic  exercise to compute the exterior derivative of \eqref{eq.faraday} to obtain the 3-form
	\begin{align}
	\label{eq.dF}
	\d F 	= 	& \sum_{i,j,k=1}^3  \frac{\partial B_{ij}}{\partial x^k}\ \d x^k \wedge \d x^i \wedge \d x^j \ +\nonumber\\
			& \sum_{\overset{i,j=1}{i\neq j}}^3 \left(\frac{\partial E_j}{\partial x^i} - \frac{\partial E_i}{\partial x^j} + \frac{\partial B_{ij}}{\partial t}  \right)\ \d x^i \wedge \d x^j \wedge \d t.
	\end{align}
It follows directly from the definition of $B_{ij}$, equation \eqref{eq.fluxcomp}, and the definition of the curl operator that the components of $\d F$ can be written as
	\begin{align}
	\d F	=	& \left( \nabla \cdot \vec B\right)\ \d x \wedge \d y \wedge \d z \ + \nonumber\\
			& \left[\left(\nabla \times \vec E + \frac{\partial \vec B}{\partial t}\right) \cdot \hat{e}_{(z)}\right] \d x \wedge \d y \wedge \d t -\nonumber\\
			& \left[\left(\nabla \times \vec E + \frac{\partial \vec B}{\partial t}\right) \cdot \hat{e}_{(y)}\right]\ \d x \wedge \d z \wedge \d t + \nonumber\\
			&\left[\left(\nabla \times \vec E + \frac{\partial \vec B}{\partial t}\right) \cdot  \hat{e}_{(x)}\right]\ \d y \wedge \d z \wedge \d t.
	\end{align}
 Thus we see that the vanishing of $\d F$ [equation \eqref{eq.MEdf2}] is completely equivalent to the the set of homogeneous Maxwell's equations. Similarly, the components of the 3-form $\d G$ corresponds to the left hand side (lhs) of the in-homogeneous Maxwell's equations \eqref{eq.EGauss} and \eqref{eq.ampere}. That is, 
 	\begin{align}
	\label{eq.dG}
	\d G 	 = 	& - \sum_{i,j,k=1}^3  \frac{\partial D_{ij}}{\partial x^k}\ \d x^k \wedge \d x^i \wedge \d x^j \ +\nonumber\\
			& \sum_{\overset{i,j=1}{i\neq j}}^3 \left(\frac{\partial H_j}{\partial x^i} - \frac{\partial H_i}{\partial x^j} - \frac{\partial D_{ij}}{\partial t}  \right)\ \d x^i \wedge \d x^j \wedge \d t,
	\end{align}
where the minus signs follow from the defintion of $G$, equation \eqref{eq.fardaydual}. Thus, subtracting the 3-form $j_{\rm ext}$, equation \eqref{eq.current}, from $\d G$ one obtains
	\begin{align}
	\d G-j_{\rm ext}=	& -\left( \nabla \cdot \vec D - \rho_{\rm ext} \right)\ \d x \wedge \d y \wedge \d z \ + \nonumber\\
			& \left[\left(\nabla \times \vec H - \frac{\partial \vec D}{\partial t} - j_{{\rm ext}}\right) \cdot \hat{e}_{(z)}\right] \d x \wedge \d y \wedge \d t -\nonumber\\
			& \left[\left(\nabla \times \vec H - \frac{\partial \vec D}{\partial t} - j_{{\rm ext}}\right) \cdot  \hat{e}_{(y)}\right]\ \d x \wedge \d z \wedge \d t + \nonumber\\
			&\left[\left(\nabla \times \vec H - \frac{\partial \vec D}{\partial t} - j_{{\rm ext}}\right) \cdot \hat{e}_{(x)}\right]\ \d y \wedge \d z \wedge \d t,
	\end{align}
whose vanishing condition \eqref{eq.MEdf2} yields the in-homogeneous Maxwell equations \eqref{eq.EGauss} and \eqref{eq.ampere}.

The exterior derivative operator is nilpotent, that is, successive applications of $\d$ are identically zero. Therefore, as before, the conservation law \eqref{eq.cont} is a consequence of the structure of Maxwell' equations, that is
	\beq
	0 = \d^2 G = \d j_{\rm ext} = \left(\frac{\partial \rho_{\rm ext}}{\partial t} + \nabla \cdot \vec j_{\rm ext} \right) \d x \wedge \d y \wedge \d z \wedge \d t.
	\eeq

Thus, we see that the differential form language appears to be tailored for electromagnetism. Moreover, equations \eqref{eq.MEdf1} and \eqref{eq.MEdf2} are not a mere abbreviation of their vectorial counterparts, as it may appear from our exercise, but a profound generalization that allows us to link the local nature of the differential equations with the global properties of their domains of definition. It is precisely this fact the one responsible for a new set of tools that has begun to be exploited in computational electromagnetism and, in particular, in the finite element method for solving electromagnetic problems in topologically complicated domains \cite{bossavit1988whitney,gross2004electromagnetic,pellikka2013homology}.

From a foundational point of view, one can reverse the argument on the conservation of total charge and take as empirical postulates the two local conservation laws 
	\beq
	\d F = 0 \quad  \text{and} \quad \d j = 0,
	\eeq
	 stating the local conservation of flux and charge, respectively. These are merely the predicates of the global postulates \eqref{eq.post1} and \eqref{eq.post2}. These imply that, at least locally in $\mathcal{M}$, there exist a pair of potentials, a 1-form $A$ and a 2-form $H$ such that 
	\beq
	F=\d A \quad \text{and}  \quad j= \d H,
	\eeq 
	where $H = G + G_{\rm ind}$, with 
	\beq 
	\label{eq.maxind}
	\d G = j_{\rm ext} \quad \text{and} \quad \d G_{\rm ind} = j_{\rm ind},
	\eeq 
	implying the independent conservation of external and induced charges. 
	Thus, the fundamental problem in electromagnetic theory can again be stated as: given a known \emph{closed} 3-form $j_{\rm ext}$, determine the \emph{closed} 2-form $F$ or, equivalently, a potential 1-form $A$. As before, this problem requires additional information linking the current density flux $j$ with the potential 1-form $A$, or the potential 2-form $G$ with the field flux 2-form $F$, namely, a constitutive relation.

\section{Geometric constitutive relations}
\label{sec.metrics}

Thus far, the differential form approach to electromagnetic theory has revealed us its topological nature. We have not introduced any information regarding its geometry, i.e. those mathematical structures that are preserved when a certain class of transformations is executed. The conservation of charge and flux are topological statements that rely solely on the differentiability of the manifold $\mathcal{M}$, not assuming any further structure. However, as we have discussed at the end of the previous section, this does not allow us to obtain the field $F$ from the given source $j_{\rm ext}$. The additional piece of information, the constitutive relation, comes at the price of demanding further structure on $\mathcal{M}$. In this manuscript, we consider the case in which such structure is given by a metric tensor 
	\beq
	g = \sum_{i,j=1}^n  g_{ij} \d x^i \otimes \d x^j
	\eeq
for   $\mathcal{M}$, i.e.  the pair $(\mathcal{M},g)$ be a (pseudo) Riemannian manifold\footnote{Pseudo Riemannian manifolds $(\mathcal{M},g)$ are those in which the metric tensor $g$  admits null vectors, that is, non-zero vectors whose norm is identically zero. In such manifolds, the Laplacian operator is hyperbolic, instead of elliptic, providing us with a suitable geometric structure to describe wave propagation.}. 

 Same as with the dot product, a metric allows one to compute lengths of parametrized curves, angles between directions at a given point and distances from one point to another  in $\mathcal{M}$ independently of the chosen coordinates. That is, these notions are \emph{invariant} under a general change of coordinates. It also serves to establish an algebraic equivalence between vectors and  1-forms by means of the \emph{musical} isomorphisms\footnote{The flat symbol $\flat$ is used to denote `lowering' the indices of the components of a vector, while the sharp symbol $\sharp$ corresponds to `raising' the indices of the components of a differential form.}, namely
  	\beq
	g^\flat(V) = \sum_{i,j=1}^n  g_{ij} V^i \d x^j \quad \text{for any} \quad V = \sum_{i=1}^n  V^i \frac{\partial}{\partial x^i},
	\eeq
and 
	\beq
	g^\sharp(\omega) = \sum_{i,j=1}^n  g^{ij} \omega_i \frac{\partial}{\partial x^j} \quad \text{for any} \quad \omega = \sum_{i=1}^n  \omega_i \d x^i.
	\eeq
In particular, for Riemannian manifolds, one is the inverse of the other, that is
	\begin{align}
	g^\sharp\left[g^\flat(V) \right] 	& = \sum_{i,j,k=1}^n  g^{ik} g_{kj} V^j \frac{\partial}{\partial x^i} \nonumber\\
							& = \sum_{i,j=1}^n  \delta^i_{\ j} V^j \frac{\partial}{\partial x^i} \nonumber\\
							& = \sum_{i=1}^n V^i \frac{\partial}{\partial x^i} \nonumber\\
							&= V,
	\end{align}
and, hence, the metric provides us with a canonical isomorphism between vector and forms.
 

  A manifold can support an infinite number of metric tensors, each one prescribing \emph{a} {geometry}. In particular,  the paths of extremal length\footnote{For a Riemannian manifold these a are the shortest paths, whilst for pseudo-Riemannian manifolds these may be the longest.}  connecting two different points in $\mathcal{M}$ may drastically differ for each pair $(\mathcal{M},g)$. In this sense, by means of \emph{Fermat's principle}, each metric tensor for $\mathcal{M}$ can be considered as a material medium for the propagation of electromagnetic waves.
 
 Let us begin by recalling the geometrization of electromagnetic theory in vacuum. To this end, consider the free space \emph{background} metric given by
 	\beq
	\label{eq.gnot}
	\eta =\sum_{i,j=1}^3  {g_0}_{ij} \d x^i \otimes \d x^j - \frac{1}{\varepsilon_0 \mu_0} \d t \otimes \d t,
	\eeq 
Here $\varepsilon_0$ and $\mu_0$ are the vacuum electric permittivity and magnetic permeability, respectively. This background metric will be assumed to correspond to the \emph{lab} space, so that the temporal basis vector
 	\beq
	u_{\rm lab} = \sqrt{\varepsilon_0 \mu_0}\ \frac{\partial}{\partial t},
	\eeq
defining the lab's rest frame, is normalized with respect to the lab metric, i.e.
 	\beq
	\label{eq.normt}
	\eta \left(u_{\rm lab}, u_{\rm lab}\right) = -1.
	\eeq

  A simple, homogeneous and isotropic medium at rest with respect to the lab frame can be characterized by a \emph{material} metric of the form
 	\beq
	\label{eq.matmet}
	g =\sum_{i,j=1}^3  g_{ij} \d x^i \otimes \d x^j - \frac{1}{\varepsilon \mu} \d t \otimes \d t.
 	\eeq
Here, $\varepsilon$ and $\mu$ are the medium's  electric permittivity and magnetic permeability, respectively, assumed to be constants. 

Notice that in the material metric, the temporal basis vector $u_{\rm lab}$ is not normalized, i.e.
	\beq
	\label{eq.normt2}
	g\left(u_{\rm lab}, u_{\rm lab}\right) = - \frac{\varepsilon_0 \mu_0}{\varepsilon \mu}.
	\eeq

 Motivated by the structure of the constitutive relations \eqref{eq.const1}, we look for a multilinear map $\kappa$ such that
	\beq
	\label{eq.consgen}
	G =\kappa[F].
	\eeq
In a Riemannian manifold, there is natural isomorphism between $p$-forms and $(n-p)$-forms associated to the metric, namely, the Hodge star operator.  Thus, let us denote $*$ the Hodge duality operator associated with the lab metric $\eta$, whilst $\star$ for the one associated with the material metric $g$. Here, we only consider its action on 2-forms. As every linear map, Hodge duality is fully defined in terms of its action on the basis forms
	\begin{align}
	* \left(\d x^i \wedge \d x^j \right)		&  = \frac{1}{\sqrt{\varepsilon_0 \mu_0}}\ \d x^k \wedge \d t,\\
	* \left(\d x^k \wedge \d t \right)   	& = -\sqrt{\varepsilon_0 \mu_0}\ \d x^i \wedge \d x^j ,\\
	\star  \left(\d x^i \wedge \d x^j \right) 	&= \frac{1}{\sqrt{\varepsilon \mu}}\ \d x^k \wedge \d t 	
	\end{align}
and
	\beq
	\star \left(\d x^k \wedge \d t \right)  	 = -\sqrt{\varepsilon \mu}\ \d x^i \wedge \d x^j.			
	\eeq

From the definition of the Hodge star operator, it is straightforward to verify that
	\begin{align}
	\star F 	& = \star B + \star \left(E \wedge \d t \right)\nonumber\\
			& = \star \left(\sum_{i,j=1}^3 B_{ij} \d x^i \wedge \d x ^j \right) + \star \left(\sum_{k=1}^3 E_k \d x^k \wedge \d t \right)\nonumber\\
			& = \sum_{i,j=1}^3 B_{ij} \star \left(\d x^i \wedge \d x ^j \right) +\sum_{k=1}^3 E_k \star \left( \d x^k \wedge \d t \right)\nonumber\\
			& = \frac{1}{\sqrt{\varepsilon \mu}}\ \sum_{k=1}^3 B_k \d x^k \wedge \d t - \sqrt{\varepsilon \mu}\  E_k \d x^i \wedge \d x^j,
	\end{align}
Therefore, the simplest constitutive relation linking the 2-forms $F$ and $G$ can be expressed in terms of the Hodge dual operator $\star$  as 
	\begin{align}
	\label{eq.hodge}
	G 	& =  \sqrt{\frac{\varepsilon}{\mu}} \star F.
	\end{align}
Indeed, cf. expressions \eqref{eq.fluxcomp},
	\begin{align}
	G	& = \frac{1}{\mu} \sum_{k=1}^3 B_k \d x^k \wedge \d t - \varepsilon E_k \d x^i \wedge \d x^j\nonumber\\
		& = \sum_{k=1}^3 H_k  \d x^k \wedge \d t - \sum_{i,j=1}^3 D_{ij}  \d x^i \wedge \d x^j\nonumber\\
		& = H \wedge \d t - D,
	\end{align}
Thus, the geometric \emph{Hodge constitutive relation} \eqref{eq.hodge} associated with the material metric \eqref{eq.matmet} is equivalent to an homogeneous and isotropic material whose constitutive relations are
 	\beq
	\vec D_{\rm lab} = \varepsilon \vec E_{\rm lab} \quad \text{and} \quad \vec H_{\rm lab} = \frac{1}{\mu} \vec B_{\rm lab}.
	\eeq
Here, the lab vector fields are obtained by contracting\footnote{The  contraction of a $p-$form and the vector field $v$ is defined as \cite{kobayashi1963foundations}
	\[
	\iota_v \omega\left[u_{(1)},\hdots,u_{(p-1)}\right] = p\cdot \omega \left[v, u_{(1)},\hdots, u_{(p-1)}\right],
	\]
where $\{u_{(i)}\}_{i=1}^{p-1}$ is a set of vector fields on $\mathcal{M}$ Thus, the contraction of a $p$-form with a vector field yields the $p-1$-form
	\[
	\iota_v \omega = p \cdot \omega(v).
	\]
} the 2-forms $F$ and $G$ with the lab frame velocity vector field $u_{\rm lab}$. Then, using the lab metric, the resulting 1-forms are mapped to their corresponding vector fields by means of its associated sharp isomorphism. That is,
%
	\begin{align}
	\label{eq.vf1}
	\vec E_{\rm lab} 	& =  - \frac{1}{\sqrt{\varepsilon_0 \mu_0}}\ \eta^\sharp\left[ \iota_{u_{\rm lab}} F\right], \\
	\vec H_{\rm lab} 	& =  - \frac{1}{\sqrt{\varepsilon_0 \mu_0}}\ \eta^\sharp\left[ \iota_{u_{\rm lab}} G\right]  
	\end{align}
and
	\begin{align}	
	\vec B_{\rm lab} 	& =  -\eta^\sharp\left[ \iota_{u_{\rm lab}}* {F} \right], \\
	\label{eq.vf4}		
	\vec D_{\rm lab} 	& =  \eta^\sharp\left[ \iota_{u_{\rm lab}} * {G} \right].
	\end{align}

Notice that, albeit \eqref{eq.vf1} - \eqref{eq.vf4} are vector fields over $\mathcal{M}$, at each tangent space these can be directly identified with the spatial vectors in $\mathbb{R}^3$ of the vector calculus formulation of electromagnetism of section \ref{sec.vector}. This conversion is usually missing in the literature of  differential forms.   

This exercise has provided us with a tool to extract  the vectorial fluxes and fields from the Faraday 2-form $F$ and a material metric $g$  in any coordinate system. Moreover, the normalized temporal vector $\frac{\partial}{\partial t}$ plays the role of an \emph{observer} at rest in the lab frame. Indeed, it is the tangent vector to a curve in $\mathcal{M}$ with no spatial components, i.e. it represents an observer spatially static moving only in the time direction at unit speed  [cf. equation  \eqref{eq.normt}]. Equations \eqref{eq.vf1} -- \eqref{eq.vf4} are the fluxes and fields \emph{seen} by a static observer in the lab frame. 
Therefore, the required closure relations for Maxwell's equations -- the constitutive relation of the medium, equation \eqref{eq.hodge} -- can be incorporated  by introducing a metric tensor representing the material. The metric is the geometry on which the electromagnetic field propagates. This feature was recognized soon after the advent of the general theory of relativity, in which a gravitational field appears as an optical medium from the point of view of light propagation. Expressing material properties in terms of \emph{curved} Riemannian manifolds is an active and fertile research area. In the present work we limit ourselves to  non-conducting, homogeneous and isotropic media. Moreover, we have seen that the observer plays a fundamental role in recovering the vectorial expressions for the fields. Indeed, the decomposition of the electromagnetic field into its electric and magnetic parts is frame dependent, i.e. different observers measure different electric and magnetic fields. 


The advantage of adopting a geometric language in formulating the constitutive relations of electromagnetism lies in its generality. Equation \eqref{eq.hodge} is observer independent and coordinate free, that is, it can be used in any coordinate system for any reference frame, inertial or not. Equations \eqref{eq.vf1} -- \eqref{eq.vf4} are expressions for the fields measured by  a static observer in the lab frame. However, they can be extended to any reference frame by replacing the static  spacetime velocity, represented by the temporal vector $\frac{\partial}{\partial t}$, by any other velocity $u$  such that  $g(u,u) = -1/\varepsilon \mu$. 

\section{The geometry of moving media}
\label{sec.movmedia}

  In this section, we will consider the effect of external electromagnetic fields on moving media. To this end, we will assume that the field $F$ is produced in the lab frame and study the \emph{induced} field $G$ in a medium described by a metric tensor adapted to the motion of an observer embedded in the material. Such motion defines a coordinate transformation  
  	\beq
	\phi:\mathcal{M} \longrightarrow \mathcal{M}
	\eeq 
mapping the material lab metric $g$ into its moving version 
	\beq 
	h = \phi^*(g)
	\eeq
which, by a fortuitous   linguistic accident, is called the \emph{induced} metric by the map $\phi$. Every geometric expression obtained in the differential form language preserves its form under such transformations.
  

Let us begin by considering two simple examples, corresponding to a Galilean and Lorentzian motions, respectively, and then we consider  non-inertial motions of the medium, namely, Galilean and relativistic rotating frames. In all cases, we consider a general electromagnetic field 2-form $F$ [cf. equations \eqref{eq.faraday}, \eqref{eq.Bform} and \eqref{eq.fluxcomp}], such that
	\beq
	\vec B_{\rm lab} 	 = 	B_x \hat{e}_{(x)} + B_y \hat{e}_{(y)} + B_z \hat{e}_{(z)} \quad \text{and} \quad \vec E_{\rm lab} 	 =  E_x \hat{e}_{(x)} + E_y \hat{e}_{(y)} + E_z \hat{e}_{(z)}.
 	\eeq

\subsection{Galilean inertially moving media}

Consider a  medium moving along the $x$ direction with constant velocity $v$. The change of coordinates associated with such a motion is given, naively, by the  Galilean transformation
	\beq
	\phi\left( 	\begin{array}{c}
			x\\
			y\\
			z\\
			t
			\end{array}
		\right)=
		\left(	\begin{array}{c}
			 x+v t\\
			 y\\
			 z\\
			 t
			 \end{array}
		\right)
%
%
	 \eeq

From the  lab's point of view, the medium is described by the material metric in the moving coordinates
	\beq
	h = \d x \otimes \d x + \d y \otimes \d y + \d z \otimes \d z + v \left( \d x \otimes \d t +\d t \otimes \d x \right)  -\frac{1}{\varepsilon\mu} \left(1 - v^2 \varepsilon \mu\right)\, \d t \otimes \d t\,.
	\eeq
Note that in these coordinates, the material metric is well defined only when
	\beq
	\label{eq.relconstr}
	v^2< \frac{1}{\varepsilon\mu},
	\eeq
that is, when the velocity of the motion is less than the speed of light in the medium.

The componentes of the vectorial electromagnetic fields induced in the moving medium as seen  by the the static observer in the lab frame [cf. equations \eqref{eq.vf1}-\eqref{eq.vf4}] are
	\begin{align}
	\label{LabF-inaive1}
	\vec D_{\rm lab}  =  \varepsilon \left( E_x \hat{e}_{(x)} + E_y \hat{e}_{(y)} + E_z \hat{e}_{(z)}  \right)
		+\varepsilon v \left( B_z \hat{e}_{(y)} - B_y \hat{e}_{(z)}\right)
	\end{align}
and
	\beq
	\label{LabF-inaive4}		
	\vec H_{\rm lab} 	 =  \frac{B_x}{\mu} \hat{e}_{(x)} + \frac{1}{\mu}\left(1-v^2 \varepsilon \mu\right) \left(  B_y \hat{e}_{(y)} + B_z \hat{e} _{(z)}\right)
	+\varepsilon v \left( E_z \hat{e}_{(y)} - E_y \hat{e}_{(z)}\right)
	\eeq
From these expressions we can read the corresponding entries of  the constitutive relations  \eqref{eq.const1}. That is,
	\beq
	\label{eq.consgal1}
	\bar{\varepsilon} = \varepsilon \left(\begin{array}{ccc}
					1 & 0 					& 0 \\
					0 & 1 	& 0 \\
					0 & 0 					&  1
					\end{array} \right),
	 \qquad
	\bar\mu^{-1} = \frac{1}{\mu} \left(\begin{array}{ccc}
					1 & 0 					& 0 \\
					0 & 1-v^2 \varepsilon \mu	& 0 \\
					0 & 0 					&  1-v^2 \varepsilon \mu
					\end{array} \right)
	\eeq
and	
	\beq
	\label{eq.consgal2}
	\bar\zeta = \bar\chi= \varepsilon v \left(\begin{array}{ccc}
					0 & 0 & 0 \\
					0 & 0 & 1 \\
					0 & -1 & 0
					\end{array} \right).
	\eeq
Observe that, for the purely electric part the medium remains homogeneous and isotropic, whilst for the magnetic field it appears to be anisotropic in the directions orthogonal to the motion.  It also appears a non-vanishing magnetoelectric matrix. Thus, from the lab point of view, when the external field is purely electric,  the induced magnetic field is perpendicular and rotating around the direction of motion. Similarly, when the externally applied field is  purely magnetic, the induced electric field has the same properties as its magnetic counterpart.  Such effect depends on the velocity of displacement of the medium with respect to the lab frame, which must satisfy \eqref{eq.relconstr}. These results are consistent with the classic results of electromagnetism in moving media, where the magnetoelectric effect is characterized by a term proportional to $\vec v\times \vec B$ for the electric part and $\vec v\times \vec E$ for the magnetic counterpart.  

%

\subsection{Lorentzian inertially moving media}

Same as in the previous example, we consider a motion along the $x$ direction, but this time by means of the transformation 
	\beq
	\phi\left( 	\begin{array}{c}
			x\\
			y\\
			z\\
			t
			\end{array}
		\right)=
		\left(	\begin{array}{c}
			 \left(1-v^2 \varepsilon_0 \mu_0\right)^{-1/2} \left( x + v t\right)\\
			 y\\
			 z\\
			  \left(1-v^2 \varepsilon_0 \mu_0\right)^{-1/2} \left(t + v x\ \varepsilon_0 \mu_0 \right)
			 \end{array}
		\right)
	\eeq
	
%
%
In this case, the material metric becomes 
	\begin{align}
	h = \left(\frac{1 - v^2\frac{\varepsilon_0^2\ \mu_0^2}{\varepsilon \mu}}{1- v^2\ \varepsilon_0 \mu_0 }\right)\ \d x \otimes \d x ++ \d y \otimes \d y + \d z \otimes \d z\nonumber \\
	+v  \left(\frac{1 - \frac{\varepsilon_0 \mu_0}{\varepsilon \mu}}{1 - v^2\ \varepsilon_0 \mu_0}\right)\ \left( \d x \otimes \d t + \d t \otimes \d x\right)\nonumber\\
	-\frac{1}{\varepsilon \mu}\left(\frac{1 - v^2 \varepsilon \mu}{1 - v^2\ \varepsilon_0 \mu_0}\right)\ \d t \otimes \d t.
	\end{align}
Again, these metric is well defined when \eqref{eq.relconstr} is satisfied. Notice that, albeit \eqref{eq.matmet} is indeed a Minkowski metric, the speed of light of the medium is, in general, different from that in vacuum. Indeed
	\beq
	\frac{\varepsilon_0 \mu_0}{\varepsilon\mu} \leq 1,
	\eeq
that is, the speed of light in the medium ought to be less than the speed of light in vacuum. Therefore, although  Lorentz transformations  leave  the vacuum metric \eqref{eq.gnot} invariant, they do change the material metric. 

  The static observer  measures the vectorial electromagnetic fields
  	\begin{align}
	\label{LabF-irel1}
	\vec D_{\rm lab} 	 =   \varepsilon E_x \hat{e}_{(x)} +  \varepsilon \left(\frac{1- v^2 \ \frac{\varepsilon_0^2 \mu_0^2}{\varepsilon \mu}}{1- v^2 \ \varepsilon_0 \mu_0} \right) \left( E_y \hat{e}_{(y)} + E_z \hat{e}_{(z)}\right)\nonumber\\ 
	+ v  \varepsilon \left(\frac{1 - \frac{\varepsilon_0 \mu_0}{\varepsilon \mu}}{1 - v^2\ \varepsilon_0 \mu_0} \right) \left(B_z\hat{e}_{(y)} -B_y\hat{e}_{(z)}\right)
	\end{align}
 %
%
and
\begin{align}
	\label{LabF-irel4}
	\vec H_{\rm lab} 	 =  \frac{1}{\mu} B_x \hat{e}_{(x)} + \frac{1}{\mu} \left(\frac{1-v^2\ \varepsilon \mu}{1-v^2\ \varepsilon_0 \mu_0}\right) \left(B_y \hat{e}_{(y)} + B_z \hat{e}_{(z)} \right) \nonumber \\
	+ v  \varepsilon \left(\frac{1 - \frac{\varepsilon_0 \mu_0}{\varepsilon \mu}}{1 - v^2\ \varepsilon_0 \mu_0} \right) \left(E_z\hat{e}_{(y)} -E_y\hat{e}_{(z)}\right)
	\end{align}
%
%
 Therefore, in  this case, the relative motion between the lab and the medium makes the material appear to the lab observer as
	\beq
	\bar{\varepsilon} = \varepsilon \left(\begin{array}{ccc}
					1 & 0 					& 0 \\
					0 & \frac{1- v^2 \ \frac{\varepsilon_0^2 \mu_0^2}{\varepsilon \mu}}{1- v^2 \ \varepsilon_0 \mu_0} 	& 0 \\
					0 & 0 					&  \frac{1- v^2 \ \frac{\varepsilon_0^2 \mu_0^2}{\varepsilon \mu}}{1- v^2 \ \varepsilon_0 \mu_0}
					\end{array} \right),
	 \qquad
	\bar\mu^{-1} = \frac{1}{\mu} \left(\begin{array}{ccc}
					1 & 0 					& 0 \\
					0 & \frac{1-v^2\ \varepsilon \mu}{1-v^2\ \varepsilon_0 \mu_0} 	& 0 \\
					0 & 0 					& \frac{1-v^2\ \varepsilon \mu}{1-v^2\ \varepsilon_0 \mu_0} 
					\end{array} \right)
	\eeq
and
	\beq
	\bar\zeta = \bar \chi= v \varepsilon  \left(\frac{1 - \frac{\varepsilon_0 \mu_0}{\varepsilon \mu}}{1 - v^2\ \varepsilon_0 \mu_0}\right) \left(\begin{array}{ccc}
					0 & 0 & 0 \\
					0 & 0 & 1 \\
					0 & -1 & 0
					\end{array} \right).
	\eeq

We see that the permitivity and permeability matrices are now anisotropic, while the magnetoelectric matrix preserves its former strucutre. Note that in the Newtoninan limit, the constitutive relations for the Galilean transformation, equations \eqref{eq.consgal1} and \eqref{eq.consgal2}, are recovered. Notice as well that in the limit when the speed of light in the medium coincides with that of vacuum, the medium becomes isotropic again and the magnetoelectric term vanishes. This shows the invariance of the vacuum with respect to Lorentz transformations.  

Hence, what it might have appeared at first glance as a simple exercise  in special relativity, it has revealed us that media in relative inertial motion acquires non-trivial electromagnetic properties as seen from another inertial frames. This does not say that the physical reality depends on the coordinates, it merely states that the constitutive relations for a simple medium in the non-covariant vector calculus lab frame are different when the medium is in relative motion.

%

\subsection{Uniformly accelerating medium}

Now we consider the medium undergoing uniform acceleration. This is the simplest  form of  non-inertial motion. Let us consider that the motion occurs along the $z$-axis with an acceleration $\alpha$, as in free fall in a uniform Newtoinian gravitational field. The transformation is written as
	\beq
	\phi\left( 	\begin{array}{c}
			x\\
			y\\
			z\\
			t
			\end{array}
		\right)=
		\left(	\begin{array}{c}
			 x\\
			 y\\
			 \left[(\alpha \varepsilon_0 \mu_0)^{-1} + z  \right] \cosh\left(\sqrt{\varepsilon_0 \mu_0}\ \alpha t \right) -(\alpha \varepsilon_0 \mu_0)^{-1}\\
			 \sqrt{\varepsilon_0 \mu_0} \left[ (\alpha \varepsilon_0 \mu_0)^{-1} + z \right] \sinh\left(\sqrt{\varepsilon_0 \mu_0}\ \alpha t \right) - (\alpha \varepsilon_0 \mu_0)^{-1}
			 \end{array}
		\right).
	\eeq
This coordinates are adapted to a uniformly accelerated observer and only cover a subset of the entire $\mathcal{M}$  referred as the Rindler wedge. 

The material metric  takes the more elaborate form
	\begin{align}
	h =\sqrt{\varepsilon_0\mu_0} \left[  \left(1 - \frac{\varepsilon_0 \mu_0}{\varepsilon \mu} \right) \alpha z - \frac{1}{\varepsilon_0 \mu_0} -  \frac{1}{\varepsilon \mu} \right] \sinh\left(\sqrt{\varepsilon_0 \mu_0}\ \alpha t \right) \cosh\left(\sqrt{\varepsilon_0 \mu_0}\ \alpha t \right) \left(\d z \otimes \d t + \d t \otimes \d z \right)\nonumber\\
	 - \left[\varepsilon_0 \mu_0 \left[\left(1 - \frac{\varepsilon_0\mu_0}{\varepsilon \mu} \right) \alpha^2 z^2 - 2 \left(\frac{1}{\varepsilon_0 \mu_0} + \frac{1}{\varepsilon\mu} \right) \alpha z  + \frac{1}{\varepsilon_0 \mu_0} \left(\frac{1}{\varepsilon_0 \mu_0} - \frac{1}{\varepsilon \mu}\right)\right] \cosh\left(\sqrt{\varepsilon_0 \mu_0}\ \alpha t \right) \right. \nonumber\\
	-\left. \frac{1}{\varepsilon_0 \mu_0} \left(1-\alpha z \varepsilon_0 \mu_0\right)^2\right] \d t \otimes \d t + \d x \otimes \d x + \d y \otimes \d y+\nonumber\\
	+ \left[ \left(1 - \frac{\varepsilon_0 \mu_0}{\varepsilon \mu} \right) \cosh^2\left(\sqrt{\varepsilon_0 \mu_0}\ \alpha t \right) + \frac{\varepsilon_0 \mu_0}{\varepsilon \mu}\right] \d z \otimes \d z,
	\end{align}
where we have the more complicated restriction
	\beq
	\cosh^2\left(\sqrt{\varepsilon_0\mu_0}\ \alpha t \right) < \left(1-\frac{\varepsilon_0 \mu_0}{\varepsilon \mu} \right)^{-1}
	\eeq
for the metric to be well defined.

The induced fields measured by the lab observer are
	\begin{align}
	\vec D_{\rm lab} =  \varepsilon \left[\left(\frac{1-\frac{\varepsilon_0\mu_0}{\varepsilon \mu}}{1+ \alpha z \varepsilon_0 \mu_0}\right) \cosh^2\left(\sqrt{\varepsilon_0 \mu_0}\ \alpha t \right)   + \frac{\varepsilon_0 \mu_0}{\varepsilon \mu} \left(\frac{1}{1+ \alpha z \varepsilon_0\mu_0}\right)\right]  \left( E_x \hat{e}_{(x)}+E_y \hat{e}_{(y)}  \right)\nonumber\\
	+ \varepsilon \left[\frac{1}{\sqrt{\varepsilon_0 \mu_0}}\left(1 - \frac{\varepsilon_0 \mu_0}{\varepsilon \mu}\right)   \sinh\left(\sqrt{\varepsilon_0 \mu_0}\ \alpha t \right)  \cosh\left(\sqrt{\varepsilon_0 \mu_0}\ \alpha t \right)  \right] \left(B_y  \hat{e}_{(x)} -B_x  \hat{e}_{(y)} \right) \nonumber\\
	+ \frac{\varepsilon E_z}{1+\alpha z \varepsilon_0 \mu_0}\ \hat{e}_{(z)}
	\end{align}
	
%
and
	\begin{align}
	\vec H_{\rm lab} = \left(\frac{1 + \alpha z \varepsilon_0\mu_0}{\mu}  \right) \left[\left(1 - \frac{\varepsilon \mu}{\varepsilon_0 \mu_0} \right)   \cosh^2\left(\sqrt{\varepsilon_0 \mu_0}\ \alpha t \right)  + \frac{\varepsilon \mu}{\varepsilon_0\mu_0} \right] \left(B_x \hat{e}_{(x)} + B_y \hat{e}_{(y)} \right)\nonumber\\
	+ \varepsilon \left[\frac{1}{\sqrt{\varepsilon_0 \mu_0}}\left(1 - \frac{\varepsilon_0 \mu_0}{\varepsilon \mu}\right)   \sinh\left(\sqrt{\varepsilon_0 \mu_0}\ \alpha t \right)  \cosh\left(\sqrt{\varepsilon_0 \mu_0}\ \alpha t \right)  \right] \left(E_y  \hat{e}_{(x)} -E_x  \hat{e}_{(y)} \right) \nonumber\\
	+ \left(\frac{1 + \alpha z \varepsilon_0\mu_0}{\mu}  \right) B_z \hat{e}_{(z)}.
	\end{align}

In this case, the constitutive relations are much more complicated. In particular, notice that the medium no longer appears to be homogeneous, there is a linear dependence on the height and, moreover, it  also seems to be time dependent. This is not surprising, since now we are \emph{measuring} the induced fields in a non-inertially moving medium from the point of view of an inertial frame.  Indeed, when the acceleration $\alpha$ is zero, we recover our original homogeneous and isotropic medium. 

The transformation considered in this section is fully consistent with special relativity. To gain some Newtonian intuition, let us consider the small acceleration limit. In this case, the induced fields take the form
	\begin{align}
	\left.\vec D_{\rm lab}\right\vert_{\alpha t \ll \frac{1}{\sqrt{\varepsilon_0^2 \mu_0^2}}} = \varepsilon \left(1 - \alpha z \varepsilon_0 \mu_0 \right)\left(E_x \hat{e}_{(x)}+E_y \hat{e}_{(y)}  + E_z \hat{e}_{(z)} \right) \nonumber\\
	 + \varepsilon  \alpha t \left(1 - \frac{\varepsilon_0 \mu_0}{\varepsilon \mu} \right) \left(B_y \hat{e}_{(x)} - B_x \hat{e}_{(y)} \right)
	\end{align}
and
	\begin{align}
	\left.\vec H_{\rm lab}\right\vert_{\alpha t \ll \frac{1}{\sqrt{\varepsilon_0^2 \mu_0^2}}} = \frac{1}{\mu} \left(1 + \alpha z \varepsilon_0\mu_0 \right)\left(B_x \hat{e}_{(x)}+B_y \hat{e}_{(y)}  + B_z \hat{e}_{(z)} \right) \nonumber\\
	+ \varepsilon  \alpha t \left(1 - \frac{\varepsilon_0 \mu_0}{\varepsilon \mu} \right) \left(E_y \hat{e}_{(x)} - E_x \hat{e}_{(y)} \right).
	\end{align}
In this limit, the medium becomes isotropic but remains inhomogeneous while the strength of the magnetoelectric effect is modulated by the ratio between the speed of light in the medium and that of the vacuum. The slower the speed of light in the medium, the greater the magnetoelectric effect. Interestingly, in the limit when $\varepsilon \mu = \varepsilon_0 \mu_0$, that is, when the moving medium is the vacuum, the medium is once again isotropic with a vanishing magnetoelectric matrix. However, it is still inhomogeneous,  i.e.
	\beq
	\left.\vec D_{\rm lab}\right\vert_{\rm vac} = \varepsilon \left(\frac{1}{1+ \alpha z \varepsilon_0\mu_0} \right) \left(E_x \hat{e}_{(x)}+E_y \hat{e}_{(y)}  + E_z \hat{e}_{(z)} \right)
	\eeq
and
	\beq
	\left.\vec H_{\rm lab}\right\vert_{\rm vac} = \frac{1}{\mu} \left(1+ \alpha z \varepsilon_0 \mu_0 \right) \left(B_x \hat{e}_{(x)}+B_y \hat{e}_{(y)}  + B_z \hat{e}_{(z)} \right).
	\eeq
This  result for the vacuum case can be read in its complementary sense, that in which the observer is the one accelerating. In such case, there is an inhomogenous apparent  polarization and magnetization of the vacuum. 

\subsection{Galilean rotating media}
\label{sec.galileanrot}

We now study  another class of non-inertially moving medium.  We shall consider a frame rotating in a Galielean fashion, followed by a rotating frame consistent with the tenets of relativity. For simplicity, let us assume that the rotation is about the $z$ axis. Therefore, in this section we work in with the metrics \eqref{eq.gnot} and \eqref{eq.matmet}, transformed into cylindrical coordinates, that is 
	\beq
	\label{eq.medcyl}
	\eta= \d r \otimes \d r + r^2 \d \varphi \otimes \d \varphi + \d z \otimes \d z - \frac{1}{\varepsilon_0 \mu_0}\d t \otimes \d t
	\eeq
and 
	\beq
	g = \d r \otimes \d r + r^2 \d \varphi \otimes \d \varphi + \d z \otimes \d z - \frac{1}{\varepsilon \mu} \d t \otimes \d t,
	\eeq
respectively.

In this coordinates, the electromagnetic 2-form is written as
	\begin{align}
	F = & E_r \d r \wedge \d t + r^2 E_\varphi \d \varphi \wedge \d t + E_z \d z \wedge \d t \nonumber \\
		&+ r\left(  B_z \d r \wedge \d \varphi -  E_\varphi \d r \wedge \d z +  E_r \d \varphi \wedge \d z\right),
	\end{align}
where 
	\begin{align}
	E_r 		& = E_y \sin(\varphi) + E_x \cos(\varphi)\\
	E_\varphi & = E_y \cos(\varphi) - E_x \sin(\varphi) 
	\end{align}
and
	\beq
	E_z = E_z,
	\eeq
while
	\begin{align}
	B_r 		& = B_x \cos(\varphi) + B_y \sin(\varphi),\\
	B_\varphi & =  B_y \cos(\varphi) - B_x \sin(\varphi),\\
	\end{align}
and
	\beq
	B_z = B_z.
	\eeq
Thus, it is straightforward to verify that
	\beq
	\vec E_{\rm lab} = E_r \hat{e}_{(r)} +  E_\varphi \frac{\hat{e}_{(\varphi)}}{r} + E_z \hat{e}_{(z)} \quad \text{and} \quad \vec B_{\rm lab} = B_r \hat{e}_{(r)} +  B_\varphi \frac{\hat{e}_{(\varphi)}}{r} + B_z \hat{e}_{(z)}.
	\eeq

 The Galilean transformation corresponding to a uniformly rotating frame with angular velocity $\omega$ is given by  
	\beq
	\phi\left( 	\begin{array}{c}
			r\\
			\varphi\\
			z\\
			t
			\end{array}
		\right)=
		\left(	\begin{array}{c}
			 r\\
			 \varphi + \omega t\\
			 z\\
			 t
			 \end{array}
		\right)
	\eeq	
 The metric for the moving medium becomes
	\begin{align}
	\label{eq.galrot}
	h=	 \d r \otimes \d r +r^2 \d \varphi \otimes \d \varphi + r^2 \omega \left( \d \varphi \otimes \d t + \d t \otimes \d \varphi \right)\nonumber\\ +\d z \otimes \d z
	 			 -\left(\frac{1}{\varepsilon \mu} - r^2 \omega^2 \right) \d t \otimes \d t\,.
	\end{align}
Now the coordinates covering $\mathcal{M}$ must satisfyy is the restriction
	\beq
	r^2 \omega^2 < \frac{1}{\varepsilon \mu}.
	\eeq
This is a constraint implying that the tangential velocity cannot be larger than the speed of light in the medium.
 
Now, the induced fields are
	\begin{align}
	\label{LabF-rnaive1}
	\vec D_{\rm lab} = \varepsilon \left(  E_r \hat{e}_{(r)} + E_\varphi \frac{\hat{e}_{(\varphi)}}{r} + E_z \hat{e}_{(z)} \right) + \varepsilon r \omega  \left(B_r \hat{e}_{(z)} - B_z \hat{e}_{(r)} \right)
	\end{align}
and
	\beq
	\label{LabF-rnaive4}		
	\vec H_{\rm lab}  =  \frac{B_\varphi}{\mu} \frac{\hat{e}_{(\varphi)}}{r} + \frac{1}{\mu}\left(1-r^2 \omega^2 \varepsilon\mu \right) \left(B_r \hat{e}_{(r)} + B_z \hat{e}_{(z)} \right) + r \varepsilon \omega \left[E_r \hat{e}_{(z)} - E_z \hat{e}_{(r)} \right].
	\eeq
%

To obtain the constitutive matrices as in the previous case, we consider the inverse cylindrical coordinates transformation. Thus, in Cartesian coordinates we have
	\beq
	\vec D_{\rm lab} = \varepsilon \left(E_x \hat{e}_{(x} + E_y \hat{e}_{(y)} + E_z \hat{e}_{(z)} \right) - \varepsilon \omega \left[ B_z x\hat{e}_{(x)} + B_z y \hat{e}_{(y)}  - \left(B_x x + B_y y \right) \hat{e}_{(z)}  \right]
	\eeq
and
	\begin{align}
	\vec H_{\rm lab} = \frac{1}{\mu}\left[ \left(1 - x^2 \omega^2 \varepsilon\mu \right) B_x \hat{e}_{(x)}+ \left(1 - y^2 \omega^2 \varepsilon\mu \right) B_y \hat{e}_{(y)}  +\left( 1- r^2 \omega^2 \varepsilon \mu\right) B_z \hat{e}_{(z)}\right]\nonumber\\
	-\varepsilon \omega \left[ E_z x\hat{e}_{(x)} + E_z y \hat{e}_{(y)}  - \left(E_x x + E_y y \right) \hat{e}_{(z)}  \right]\nonumber\\
	-xy \omega^2\varepsilon  \left(   B_y \hat{e}_{(x)} +   B_x \hat{e}_{(y)}  \right)
	\end{align}
Therefore, the constitutive relations are expressed as
	\beq
	\bar{\varepsilon} = \varepsilon \left(\begin{array}{ccc}
					1 & 0 					& 0 \\
					0 & 1 	& 0 \\
					0 & 0 					&  1
					\end{array} \right),
	\eeq
	\beq
		\bar\mu^{-1} = \frac{1}{\mu} \left(\begin{array}{ccc}
					1 - x^2 \omega^2 \varepsilon\mu  & -xy \omega^2\varepsilon\mu  					& 0 \\
					-xy \omega^2\varepsilon\mu  & 1 - y^2 \omega^2 \varepsilon\mu  	& 0 \\
					0 & 0 					&  1- r^2 \omega^2 \varepsilon \mu
					\end{array} \right),
	\eeq
while, the magnetoelectric matrix is given by
	\beq
	\label{eq.magelrot}
	\bar\zeta=\bar\chi  = -\varepsilon \omega  \left(\begin{array}{ccc}
					0 & 0 & x \\
					0 & 0 & y \\
					-x & -y & 0
					\end{array} \right).
	\eeq

This constitutive matrices describe a trivial permittivity  but a much more complex permeability which, in this case,  is inhomogeneous and anisotropic. This, however, is only noticeable far from the axis of rotation, when the tangential velocity approaches the speed of light in the medium. However, note that the magnetoelectric matrix in non-negligible for any angular velocity.

%
\subsection{Relativistic rotating media}

Considering the material rotating as before, but now, we will transform the coordinates of the moving medium taking into account special relativity for the rotation  \cite{AshworthDG1979}. In this case, for a given angular velocity $\omega$, there is a maximum distance $R$ to the axis of rotation. This corresponds to the upper bound  for the radial coordinate such that the norm of the tangential velocity is less than the speed of light in vacuum. Here, $R$ is a metric parameter. Each value of $R$ and $\omega$ yield a different metric. These coordinates only cover a region of Minkowski spacetime and there is a horizon for each value of $R$ and $\omega$. Thus, let us consider the transformation
	\beq
	\phi\left( 	\begin{array}{c}
			r\\
			\varphi\\
			z\\
			t
			\end{array}
		\right)=
		\left(	\begin{array}{c}
			r  (1-R^2\omega^2 \varepsilon_0 \mu_0)^{\frac{1}{2}}\\
			 (\varphi - \omega t)(1-R^2\omega^2 \varepsilon_0 \mu_0)^{-\frac{1}{2}}\\
			 z\\
			 t  (1-R^2\omega^2 \varepsilon_0 \mu_0)^{\frac{1}{2}}
			 \end{array}
		\right)
	\eeq	
As many authors have noted, this is not the only possibility for describing a rotating reference frame. This is indeed a timely problem and there are numerous presentations of the paradoxes and issues associated with relativistic  rotating frames.

The induced material metric  takes the form
	\begin{align}
	\label{eq.rotmetR}
	h=	\left(1+ R^2\omega^2\varepsilon_0\mu_0\right) \ \d r \otimes \d r +r^2  \d \varphi \otimes \d \varphi + r^2 \omega\left(\d \varphi \otimes \d r + \d t \otimes \d \varphi\right) \nonumber\\
	  + \d z \otimes \d z  - \left[\frac{1}{\varepsilon\mu} - R^2\omega^2\left(  \frac{r^2}{R^2} -\frac{\varepsilon_0\mu_0}{\varepsilon\mu}  \right)  \right] \, \d t \otimes \d t.
	\end{align}
Note that this metric has a richer structure than our previous example. For instance, the parameters $R$ and $\omega$ must  satisfy the restriction that that the tangential velocity never exceeds that of light in vacuum, that is
	\beq
	R^2\omega^2 < \frac{1}{ \varepsilon_0\mu_0}.
	\eeq
In addition,  we can see that  these coordinates only cover the region where
	\beq
	r^2 \omega^2 < \frac{1}{\varepsilon \mu} - R^2 \omega^2\left(\frac{\varepsilon_0\mu_0}{\varepsilon\mu}\right).
	\eeq
Such bound can be regarded as the maximum tangential speed the material can attain. Moreover, note that in the limit where the tangential velocity $R \omega$ coincides with the speed of light in the vacuum, the region degenerates to a point. However, in the non-relativistic limit, namely, when $R^2 \omega^2 \ll 1/\sqrt{\varepsilon_0\mu_0}$,  \eqref{eq.rotmetR} reduces to the Galilean rotating metric \eqref{eq.rotmetR}.  Finally, as expected, in the limit when $\omega$ vanishes we return to the static metric \eqref{eq.medcyl}. 

The vectorial induced electromagnetic fields measured in the lab frame are
	\begin{align}
	\label{LabF-rrel1}
	\vec D_{\rm lab}  = \varepsilon \left[\left( \frac{1}{1 + R^2 \omega^2 \varepsilon_0\mu_0}\right)   E_r \hat{e}_{(r)}  + E_\varphi \frac{\hat{e}_{(\varphi)}}{r}+ E_z \hat{e}_{(z)} \right]\nonumber\\
		+ \varepsilon \omega r \left[ B_r \hat{e}_{(z)} -  \left( \frac{1}{1 + R^2 \omega^2 \varepsilon_0\mu_0}\right)   B_z \hat{e}_{(r)} \right]
	\end{align}
and
	\begin{align}
	\label{LabF-rrel4}		
	\vec H_{\rm lab} =  \frac{1}{\mu} \left[1 - R^2\omega^2 \varepsilon\mu\left(  \frac{r^2}{R^2} -\frac{\varepsilon_0\mu_0}{\varepsilon\mu}  \right)  \right]\left( B_r \hat{e}_{(r)} + B_z \hat{e}_{(z)} \right)  +\frac{ B_\varphi}{\mu}\ \frac{\hat{e}_{(\varphi)}}{r}\nonumber\\
	+ \varepsilon \omega r \left[ \left( \frac{1}{1 + R^2 \omega^2 \varepsilon_0\mu_0}\right) E_r \hat{e}_{(z)} -     E_z \hat{e}_{(r)} \right].
	\end{align}
	
Again, it is not difficult to express these fields in Cartesian coordinates
	\begin{align}
	\vec D_{\rm lab}  = \left(\frac{\varepsilon}{1 + R^2 \omega^2 \varepsilon_0\mu_0} \right)\left[\left(1 + \frac{R^2}{r^2}  \omega^2 y^2 \epsilon_0 \mu_0\right) E_x \hat{e}_{(x)} +\left(1 + \frac{R^2}{r^2}  \omega^2 x^2 \epsilon_0 \mu_0\right) E_y \hat{e}_{(x)}   \right] \nonumber\\
	+ \varepsilon  E_z \hat{e}_{(z)}- \varepsilon\omega^2 xy  \left(\frac{R^2  \varepsilon_0\mu_0}{r^2 \left(1 + R^2 \omega^2 \varepsilon_0\mu_0 \right)}  \right)\left( E_y \hat{e}_{(x)} +E_x \hat{e}_{(y)} \right)\nonumber\\
	- \varepsilon  \omega\left[ \left(\frac{1}{1 + R^2 \omega^2 \varepsilon_0\mu_0} \right) \left(x B_z \hat{e}_{(x)} +  y B_z  \hat{e}_{(y)}\right) - \left(x B_x + y B_y \right) \hat{e}_{(z)}\right] 
	\end{align}
and
	\begin{align}
	\vec H_{\rm lab} = \frac{1}{r^2\mu}\left[\omega^2 \varepsilon\mu\left( x^2y^2 - x^4 \right)    + \left(1 + R^2 \omega^2 \varepsilon_0\mu_0\right) x^2\right] B_x \hat{e}_{(x)}\nonumber\\
	+ \frac{1}{r^2\mu}\left[\omega^2 \varepsilon\mu\left( x^2y^2 - y^4 \right)    + \left(1 + R^2 \omega^2 \varepsilon_0\mu_0\right) y^2\right] B_y \hat{e}_{(y)}\nonumber\\
	+ \frac{1}{\mu}\left(\frac{1}{1 + R^2 \omega^2 \varepsilon_0\mu_0} \right)\left[1-r^2\varepsilon\mu \left(1 - \frac{R^2}{r^2} \frac{\varepsilon_0\mu_0}{\varepsilon \mu} \right)  \right] B_z \hat{e}_{(z)}\nonumber\\
	-   \varepsilon \omega^2 x y \left(1- \frac{R^2}{r^2} \frac{\varepsilon_0\mu_0}{\varepsilon\mu} \right) \left( B_y \hat{e}_{(x)}+B_x \hat{e}_{(y)} \right)\nonumber\\
	 - \varepsilon \omega\left[\left(x E_z\hat{e}_{(x)} +y E_z\hat{e}_{(y)}\right) -\left(\frac{1}{1 + R^2 \omega^2 \varepsilon_0\mu_0} \right) \left(x E_x + y E_y \right) \hat{e}_{(z)} \right].
	\end{align}

This frame yields a highly non-trivial material medium as seen from the lab frame. In particular, note that in all cases the behavior of the magnetic part is significantly different from the electric one. Furthermore, this example shows that the magnetoelectric matrices can differ. Indeed
	\beq
	\bar\zeta  = -\varepsilon \omega  \left(\begin{array}{ccc}
					0 & 0 &  \frac{x}{1 + R^2 \omega^2 \varepsilon_0\mu_0} \\
					0 & 0 &  \frac{y}{1 + R^2 \omega^2 \varepsilon_0\mu_0} \\
					-x & -y & 0
					\end{array} \right)
	\eeq
whilst
	\beq
	\bar\chi  = -\varepsilon \omega  \left(\begin{array}{ccc}
					0 & 0 &  x \\
					0 & 0 & y \\
					 \frac{-x}{1 + R^2 \omega^2 \varepsilon_0\mu_0} & \frac{-y}{1 + R^2 \omega^2 \varepsilon_0\mu_0} & 0
					\end{array} \right).
	\eeq

%
%
%

Hence, the lesson this exercise exhibits is that, while the medium at rest can indeed be as simple as possible, its motion renders a more complicated material structure. That is, we can think of the moving material as an  \emph{equivalent medium} at rest in the lab frame but with a  much more elaborate constitutive relation. Moreover, the calculations are simple contractions and canonical mappings between differential forms and vector fields, showing the power of the geometric formalism in obtaining  the non-covariant components of the induced fields in $\mathbb{R}^3$ along with their constitutive matrices.

\section{A non-trivial medium}
\label{sec.nontrivial}

For completeness, we explore a non-trivial medium, i.e. one whose metric yields a non-zero curvature and which has been studied in the context of transformation optics and analogue gravity \cite{AnalogueGravPhen}.  Let us consider a fisheye lens, an optical medium whose geometry is equivalent to that of the Einstein universe
	\beq
	\label{fisheye.metric}
	g_{\rm \Lambda} = \d x \otimes \d x + \d y \otimes \d y + \d z \otimes \d z - \frac{1}{\varepsilon \mu} \left[\frac{k \Lambda}{16} \left(\frac{4}{k} + x^2 + y^2 + z^2 \right)^2 \right] \d t \otimes \d t
	\eeq 
where $k$ is a constant representing the Gaussian curvature of the space and $\Lambda$ is the cosmological constant \cite{ExactSol-Stephani}. Note that this geometry is not flat, as in the previous cases. In particular, its curvature is completely specified by its Ricci scalar
	\beq
	S =- \frac{12}{\frac{4}{k} + x^2 + y^2 + z^2}.
	\eeq
	
The metric \eqref{fisheye.metric}  is written in isotropic cartesian coordinates, allowing us to read directly  the effective velocity of light in the medium [cf. equation \eqref{eq.matmet}]. In particular, the refractive index is
	\beq
	\label{refractive.index}
	n^2= \frac{9 k \Lambda}{ S^2} \frac{\varepsilon\mu}{\varepsilon_0 \mu_0}.
	\eeq

In this case, the geometry corresponds to a non-homogeneous medium, as can be directly verified by the corresponding rest frame induced fields
	\beq
	\vec D_{\rm lab} = \frac{4 \varepsilon}{\left(\frac{4}{k} + x^2 + y^2 +z^2 \right) \sqrt{k \Lambda}} \vec E_{\rm lab} =-\frac{1}{3} \frac{S}{\sqrt{k\Lambda}} \varepsilon \ \vec E_{\rm lab}
	\eeq
and
	\beq
	\vec H_{\rm lab} =\frac{\left(\frac{4}{k} + x^2 + y^2 +z^2 \right)\sqrt{k \Lambda}}{4  \mu} \vec B_{\rm lab} =-\frac{3}{\mu}  \frac{\sqrt{k \Lambda}}{S}  \ \vec B_{\rm lab}. 
	\eeq
Thus, the  lab frame permitivity and permeability matrices are
	\beq
	\label{eq.fisheyerest}
	\bar{\varepsilon} =-\frac{1}{3} \frac{S}{\sqrt{k \Lambda}} \varepsilon \left(\begin{array}{ccc}
					1 & 0 					& 0 \\
					0 & 1 	& 0 \\
					0 & 0 					&  1
					\end{array} \right)
	 \quad \text{and} \quad
	\bar\mu^{-1} =- \frac{3}{\mu} \frac{\sqrt{k\Lambda}}{S}  \left(\begin{array}{ccc}
					1 & 0 					& 0 \\
					0 & 1	& 0 \\
					0 & 0 					&  1
					\end{array} \right).
	\eeq

The transformations studied throughout the manuscript render the algebraic expressions for the induced metric and fields rather cumbersome with little conceptual value. It is a mere exercise in differential geometry to obtain them. Moreover, since the expressions presented here are coordinate free, we are guaranteed that all the coordinate expressions are indeed self consistent within the formalism. Nonetheless, let us consider the Galilean transformation to a  rotating frame presented in section \ref{sec.galileanrot}. In this frame, the induced fields as seen by the lab observer become	
	\begin{align}
	\vec D_{\rm lab} = &	 -\frac{1}{3} \frac{S}{\sqrt{k\Lambda}} \varepsilon\left[ E_x \hat e_{(x)} + E_y \hat e_{(y)} + E_z \hat e_{(z)} \right]\nonumber\\
	 & -\frac{1}{3} \frac{S}{\sqrt{k \Lambda}} \omega \varepsilon \left[-B_z \left( x \hat e_{(x)} +  y \hat e_{(y)}\right) +  \left(B_x x + B_y y \right) \hat e_{(z)} \right]
	\end{align}
and
	\begin{align}
	\vec H_{\rm lab} = &  - \frac{3}{\mu} \frac{\sqrt{k \Lambda}}{S}  \left[\left(1-\frac{ S^2}{9 k \Lambda} \omega^2  \varepsilon\mu x^2  + \frac{ S^2}{9 k \Lambda} \omega^2 (x^2 + y^2) \varepsilon\mu\right) B_y  - \left(\frac{ S^2}{9 k \Lambda} \omega^2  \varepsilon\mu xy\right) B_x \right]\hat e_{(y)}
 \nonumber\\
 		 &- \frac{3}{\mu}  \frac{\sqrt{k \Lambda}}{S} \left[\left(1-\frac{ S^2}{9 k \Lambda} \omega^2  \varepsilon\mu x^2  \right) B_x - \left(\frac{ S^2}{9 k \Lambda} \omega^2  \varepsilon\mu xy\right) B_y  \right] \hat e_{(x)}\nonumber\\
		  &- \frac{3}{\mu}  \frac{\sqrt{k \Lambda}}{S} \left[1 -\frac{ S^2}{9 k \Lambda} \omega^2 (x^2+y^2) \varepsilon\mu \right] B_z \hat e_{(z)}\nonumber\\
		  & -\frac{1}{3} \frac{S}{\sqrt{k \Lambda}} \omega \varepsilon  \left[ -E_z \left(x \hat e_{(x)} +y \hat e_{(y)} \right)  + \left(E_x x + E_y y \right)\hat e_{(z)} \right] .
		\end{align}
	
Again, we can simply read the corresponding constitutive matrices for the moving medium as seen in the lab frame. Thus, we see that the permitivity matrix remains the same as in \eqref{eq.fisheyerest}, while the permeability matrix becomes
	\beq
		\bar\mu^{-1} =- \frac{3}{\mu} \frac{\sqrt{k \Lambda}}{S}  \left(\begin{array}{ccc}
					1-\frac{ S^2}{9 k \Lambda} \omega^2  \varepsilon\mu x^2 & \frac{S^2}{9 k\Lambda} \omega^2  \varepsilon\mu xy 					& 0 \\
					\frac{ S^2}{9 k \Lambda} \omega^2  \varepsilon\mu xy & 1-\frac{ S^2}{9 k \Lambda} \omega^2  \varepsilon\mu x^2  + \frac{ S^2}{9 k \Lambda} \omega^2 (x^2 + y^2) \varepsilon\mu	& 0 \\
					0 & 0 					&  1 -\frac{ S^2}{9 k \Lambda} \omega^2 (x^2+y^2) \varepsilon\mu
					\end{array} \right).
	\eeq
Note that the non-trivial terms in this expression are quadratic in the angular velocity. Therefore, for small tangential velocities  compared with the speed of light in the medium the permeability reduces to that in \eqref{eq.fisheyerest}. 

Finally, similar to equation \eqref{eq.magelrot} in \ref{sec.galileanrot}, the magentoelectric matrices are
	\beq
	\bar\zeta=\bar\chi  =  -\frac{1}{3} \frac{S}{\sqrt{k \Lambda}} \omega \varepsilon  \left(\begin{array}{ccc}
					0 & 0 & -x \\
					0 & 0 & -y \\
					x & y & 0
					\end{array} \right).
	\eeq

In this last example, we obtained -- as expected -- similar results as those of \ref{sec.galileanrot}. However, note that the curvature plays a central role in the constitutive relations. Therefore, this exercise allows us to see the effectiveness of the formalism in obtaining explicit coordinate expressions for the constitutive matrices of a medium moving in an arbitrary reference frame. Finally, note that for the slow velocity regime the refractive index remains unchanged, in spite   the magnetoelectric matrices are not negligible. 

\section{Closing remarks}
\label{sec.concl}

In this work, it was our aim to present  to a broader readership the geometric techniques in electromagnetic theory. In particular, we addressed a subtle and timely subject, that is, the transformation of the constitutive relations for arbitrarily moving media. We considered the case of familiar motions in both, the more intuitive Galilean framework and the one consistent with the tenets of special relativity, whose symmetry is precisely that stemming from electromagnetism. 

We began with a brief summary of college electromagnetism followed by its modern formulation in terms of differential forms. We noted that  Maxwell's  empirical postulates are of topological nature on a differentiable manifold. There is no need of an additional geometric structure. However, as a field theory problem, i.e. determining the fields from the known external sources, we need a link between the two postulates [cf. equations \eqref{eq.post1} and \eqref{eq.post2}]. In the simplest case, such a link is linear. It has been argued that it may appear as a curvature-like tensor \cite{hehl2012foundations,asenjo2017differential}. Such approach, is more general than the metric based  considerations followed in this manuscript. Nevertheless, with no canonical way of mapping differential forms to vector fields, it is conceptually harder and there would be no natural way to recover the vectorial components of the electromagnetic fields. Thus, in this manuscript, we postulated the constitutive relations through the Hodge duality associated with the metric characterizing the medium \cite{baldomir1996geometry}.  We provided explicit formulae for the spatial vector fields (in $\mathbb{R}^3$) measured by the lab observer. This connection with the old fashioned --yet widely used -- vector calculus formulation of electromagnetism in media is, to the best of our knowledge, not widely known. Moreover, the calculation is coordinate independent and can be adapted to an arbitrary observer.

We used the expressions of the lab frame spatial vector fields, equations \eqref{eq.vf1} - \eqref{eq.vf4}, to compute the induced fields in a homogeneous and isotropic medium when it is set in distinct types of motion. Such motions are given in terms of  changes of coordinates acting on the material metric, alone. The induced fields $G$ are computed by applying the Hodge constitutive relation of the transformed metric to the untransformed external 2-form $F$, equation \eqref{eq.hodge}, and then contracting the result with the lab frame velocity and using the lab metric sharp isomorphism to obtain the desired vectors. A similar `mixed' approach for the vector calculus formulation can be found in Section 9-5 of \cite{panofsky2005classical}. 

As it may appear that other efforts have been successful in describing the electromagnetic fields when the medium is in motion \cite{RousseauxG06-OEMBLV,RousseauxG08-OEMLV}, we tackled a different problem. In this work, we exhibited the  explicit form of the induced electromagnetic fields \emph{measured} by a static observer when the medium moves in an arbitrary fashion. Moreover, we recovered the coordinate expressions for the permittivity, permeability and magnetoelectric matrices for non-inertial motions even for non-trivial media. 

In the case of the Galilean inertially moving media, for the purely electric part, the permittivity of the medium remains  homogeneous and isotropic while there is also  an induced  magnetic field rotating around the direction of the motion and whose magnitude depends on the velocity of displacement of the medium. This is the magnetoelectric effect and is expressed as a non-vanishing magnetoelectric matrix [cf. equation \eqref{eq.const1}]. In contrast,  the purely magnetic field  generates an anisotropic permeability matrix and,  similarly to the electric case, a rotating induced electric field is obtained. 
For the Lorentzian transformation of coordinates, both matrices, permittivity and permeability became anisotropic, while the magnetoelectric matrix is merely a rescaling from its Galilean counterpart. In the limit when the speed of light in the medium coincides with the one in vacuum, the medium returns to be isotropic and the megnetoelectric matrix vanishes, showing the invariance of the vacuum with respect to Lorentz transformations. 

We also considered a medium undergoing uniform acceleration. This  resulted in a material which is inhomogeneous, anisotropic and time-dependent. This showed us that, even in the simplest form of non-inertial motion, the medium becomes already very complex from the point of view of an inertial frame of reference.

In the case of a rotating  medium, for the Galilean-like transformation, the permittivity matrix remained the same as in the static lab frame. However, as in the slow acceleration case, the permeability matrix is inhomogeneous and anisotropic. The  magnetoelectric matrix  is also inhomogeneous  and its effects can be observed for any angular velocity $\omega$. If we also take into account special relativity in the definition of the transformation, the rotating medium yields a highly non-trivial equivalent material  as seen from the lab's rest frame. 


Hence, this work presents an algebraic method to obtain the constitutive matrices for moving media as measured by an  inertial observer. In particular, this tool provides us, in a completely covariant manner, with a way to compute the induced vector fields on such media. Moreover, this same methodology can be applied to more complicated materials -- those described by  curved geometries --  in arbitrary motion without further modification. 

\section*{Acknowledgment}

DGP is  funded by a CONACYT PhD Scholarship CVU 425313. 
\section*{References}

 \bibliography{biblio}

\begin{thebibliography}{10}
\expandafter\ifx\csname url\endcsname\relax
  \def\url#1{\texttt{#1}}\fi
\expandafter\ifx\csname urlprefix\endcsname\relax\def\urlprefix{URL }\fi
\expandafter\ifx\csname href\endcsname\relax
  \def\href#1#2{#2} \def\path#1{#1}\fi

\bibitem{gordon1923lichtfortpflanzung}
W.~Gordon, Zur lichtfortpflanzung nach der relativit{\"a}tstheorie, Annalen der
  Physik 377~(22) (1923) 421--456.

\bibitem{EhlersJ2012GFFandLP}
J.~Ehlers, F.~Pirani, A.~Schild, The geometry of free fall and light
  propagation, General Relativity and Gravitation 44 (2012) 1587--1609.

\bibitem{de1971gravitational}
F.~de~Felice, On the gravitational field acting as an optical medium, General
  Relativity and Gravitation 2~(4) (1971) 347--357.

\bibitem{de2002analogue}
V.~De~Lorenci, R.~Klippert, Analogue gravity from electrodynamics in nonlinear
  media, Physical Review D 65~(6) (2002) 064027.

\bibitem{novello2003analogue}
M.~Novello, S.~P. Bergliaffa, J.~Salim, V.~De~Lorenci, R.~Klippert, Analogue
  black holes in flowing dielectrics, Classical and Quantum Gravity 20~(5)
  (2003) 859.

\bibitem{belgiorno2011dielectric}
F.~Belgiorno, S.~Cacciatori, G.~Ortenzi, L.~Rizzi, V.~Gorini, D.~Faccio,
  Dielectric black holes induced by a refractive index perturbation and the
  hawking effect, Physical Review D 83~(2) (2011) 024015.

\bibitem{hehl2012foundations}
F.~W. Hehl, Y.~N. Obukhov, Foundations of classical electrodynamics: Charge,
  flux, and metric, Vol.~33, Springer Science \& Business Media, 2012.

\bibitem{asenjo2017differential}
F.~A. Asenjo, C.~Erices, A.~Gomberoff, S.~A. Hojman, A.~Montecinos,
  Differential geometry approach to asymmetric transmission of light, Optics
  express 25~(22) (2017) 26405--26416.

\bibitem{leonhardt2006general}
U.~Leonhardt, T.~G. Philbin, General relativity in electrical engineering, New
  Journal of Physics 8~(10) (2006) 247.

\bibitem{zheludev2010road}
N.~I. Zheludev, The road ahead for metamaterials, Science 328~(5978) (2010)
  582--583.

\bibitem{schuster2017effective}
S.~Schuster, M.~Visser, Effective metrics and a fully covariant description of
  constitutive tensors in electrodynamics, Physical Review D 96~(12) (2017)
  124019.

\bibitem{thompson2018covariant}
R.~T. Thompson, Covariant electrodynamics in linear media: Optical metric,
  Physical Review D 97~(6) (2018) 065001.

\bibitem{schuster2019electromagnetic}
S.~Schuster, M.~Visser, Electromagnetic analogue space-times, analytically and
  algebraically, Classical and Quantum Gravity.

\bibitem{PendryJ06-CEF}
D.~R.~S. J.~B.~Pendry, D.~Schurig, Controlling electromagnetic fields, Science
  312 (2006) 1780-- 1782.

\bibitem{misner1957classical}
C.~W. Misner, J.~A. Wheeler, Classical physics as geometry, Annals of physics
  2~(6) (1957) 525--603.

\bibitem{misner1973gravitation}
C.~W. Misner, K.~S. Thorne, J.~A. Wheeler, et~al., Gravitation, Macmillan,
  1973.

\bibitem{baldomir1996geometry}
D.~Baldomir, P.~Hammond, Geometry of electromagnetic systems, no.~39, Oxford
  University Press, 1996.

\bibitem{bossavit1999computational}
A.~Bossavit, {\'E}.~́~De~France, Computational electromagnetism and geometry.

\bibitem{leonhardt2010geometry}
U.~Leonhardt, T.~Philbin, Geometry and light: the science of invisibility,
  Courier Corporation, 2010.

\bibitem{stenvall2013manifolds}
A.~Stenvall, T.~Tarhasaari, F.~Grilli, P.~Raumonen, M.~Vojen{\v{c}}iak,
  M.~Pellikka, Manifolds in electromagnetism and superconductor modelling:
  Using their properties to model critical current of twisted conductors in
  self-field with 2-d model, Cryogenics 53 (2013) 135--141.

\bibitem{gron2004space}
{\O}.~Gr{\o}n, Space geometry in rotating reference frames: A historical
  appraisal, in: Relativity in Rotating Frames, Springer, 2004, pp. 285--333.

\bibitem{rizzi2013relativity}
G.~Rizzi, M.~L. Ruggiero, Relativity in Rotating Frames: Relativistic Physics
  in Rotating Reference Frames, Vol. 135, Springer Science \& Business Media,
  2013.

\bibitem{gourgoulhon2016special}
{\'E}.~Gourgoulhon, Special relativity in general frames, Springer, 2016.

\bibitem{rontgen1888ueber}
W.~C. R{\"o}ntgen, Ueber die durch bewegung eines im homogenen electrischen
  felde befindlichen dielectricums hervorgerufene electrodynamische kraft,
  Annalen der Physik 271~(10) (1888) 264--270.

\bibitem{fiebig2005revival}
M.~Fiebig, Revival of the magnetoelectric effect, Journal of physics D: applied
  physics 38~(8) (2005) R123.

\bibitem{spaldin2019advances}
N.~A. Spaldin, R.~Ramesh, Advances in magnetoelectric multiferroics, Nature
  materials 18~(3) (2019) 203.

\bibitem{DahlM04-CGE}
M.~Dahl, Contact geometry in electromagnetism, Progress In Electromagnetics
  Research 46 (2004) 77--104.

\bibitem{nash1988topology}
C.~Nash, S.~Sen, Topology and geometry for physicists, Elsevier, 1988.

\bibitem{nakahara2003geometry}
M.~Nakahara, Geometry, topology and physics, CRC Press, 2003.

\bibitem{kobayashi1963foundations}
S.~Kobayashi, K.~Nomizu, Foundations of differential geometry, Vol.~1, New
  York, London, 1963.

\bibitem{gross2004electromagnetic}
P.~W. Gross, P.~W. Gross, P.~R. Kotiuga, R.~P. Kotiuga, Electromagnetic theory
  and computation: a topological approach, Vol.~48, Cambridge University Press,
  2004.

\bibitem{WarnickK-2014}
K.~F. Warnick, P.~Russer, Differential forms and electromagnetic field theory,
  Progress In Electromagnetics Research 148 (2014) 83--112.

\bibitem{bossavit1988whitney}
A.~Bossavit, Whitney forms: A class of finite elements for three-dimensional
  computations in electromagnetism, IEE Proceedings A (Physical Science,
  Measurement and Instrumentation, Management and Education, Reviews) 135~(8)
  (1988) 493--500.

\bibitem{pellikka2013homology}
M.~Pellikka, S.~Suuriniemi, L.~Kettunen, C.~Geuzaine, Homology and cohomology
  computation in finite element modeling, SIAM Journal on Scientific Computing
  35~(5) (2013) B1195--B1214.

\bibitem{AshworthDG1979}
D.~G. Ashworth, P.~A. Davis, Transformations between inertial and rotating
  frames of reference, J. Phys. A: Math. Gen. 12~(9).

\bibitem{ExactSol-Stephani}
H.~Stephani, D.~Kramer, M.~MacCallum, C.~Hoenselaers, E.~Herlt, Exact Solutions
  of Einstein's Field Equations, 2nd Edition, Cambridge University Press, 2003.

\bibitem{AnalogueGravPhen}
D.~Faccio, F.~Belgiorno, et. al. (Eds.), Analogue Gravity Phenomenology, Vol.
  870, Springer.

\bibitem{panofsky2005classical}
W.~K. Panofsky, M.~Phillips, Classical electricity and magnetism, Courier
  Corporation, 2005.

\bibitem{RousseauxG06-OEMBLV}
M.~de~Montigny, G.~Rousseaux, On the electrodynamics of moving bodies at low
  velocities, European Journal of Physics 27 (2006) 755--768.

\bibitem{RousseauxG08-OEMLV}
G.~Rousseaux, On the electrodynamics of minkowski at low velocities, EPL 84.

\end{thebibliography}

\end{document}